\documentclass[final]{article}
\usepackage{spconf,amsmath,hyperref}
\usepackage{booktabs}
\usepackage{siunitx}
\usepackage{graphicx}
\usepackage{amsmath}
\usepackage{amsfonts}
\usepackage{float}

\newcommand{\amy}[1]{\textcolor{black}{#1}}

\title{{Diff-VS}: Efficient Audio-Aware Diffusion U-Net for Vocals Separation}

\name {Yun-Ning (Amy) Hung, Richard Vogl, Filip Korzeniowski, Igor Pereira}
\address{Moises, USA}

\begin{document}
%
\maketitle
\begin{abstract}
While diffusion models are best known for their performance in generative tasks, they have also been successfully applied to many other tasks, including audio source separation. However, current generative approaches to music source separation often underperform on standard objective metrics. In this paper, we address this issue by introducing a novel generative vocal separation model based on the Elucidated Diffusion Model (EDM) framework. Our model processes complex short-time Fourier transform spectrograms and employs an improved U-Net architecture based on music-informed design choices. Our approach matches discriminative baselines on objective metrics and achieves perceptual quality comparable to state‑of‑the‑art systems, as assessed by proxy subjective metrics. We hope these results encourage broader exploration of generative methods for music source separation.
\end{abstract}
\begin{keywords}
Diffusion Model, Music Source Separation, Vocal Source Separation
\end{keywords}
\section{Introduction}
\label{sec:intro}

Recent advances in music source separation using discriminative approaches, where models focus on directly mapping the input mixture to its separated source, have shown promising results. Leveraging band-splitting and RoFormer-based architectures, BS-RoFormer \cite{lu2024music} and Mel-RoFormer \cite{wang2024mel} demonstrate strong performance and good scalability on four-stem separation. With band-splitting and a sparse compression network, SCNet \cite{tong2024scnet} achieves competitive performance with faster inference, fewer parameters, and a single model for all four stems. Moreover, Hung et~al.\ \cite{moises-light} show that, with appropriate model design and an efficient band-splitting mechanism, compact U-Net architectures can achieve results comparable to much larger counterparts.

Generative approaches, on the other hand, have been less explored for music source separation, despite their successes in related tasks such as audio generation \cite{evans2025stable} and speech enhancement \cite{welker22speech}. Mariani et~al.\ \cite{mariani2023multi} and Karchkhadze et~al.\ \cite{karchkhadze2025simultaneous} unify separation and generation within a single framework. Although they demonstrate basic separation capabilities, the models require 150 sampling steps during inference and comprise over 400M parameters. Furthermore, since they were trained exclusively on synthesized instrumental music, the generalization capabilities to real-world music are not well studied. More recent work \cite{plaja2025diff} investigates Denoising Diffusion Probabilistic Models (DDPM) within the UNIVERSE \cite{serra2022universal} diffusion framework. This approach processes plain waveforms for music source separation, reducing the model size to 99M parameters and the number of required inference steps to roughly 20. To further accelerate inference, Plaja-Roglans et~al.\ \cite{plaja2025latentdiff} apply diffusion to latent embeddings from EnCodec \cite{defossez2022highfi}. Other hybrid approaches, such as \cite{stoller2018adversarial} and \cite{bereuter2025towards}, combine generative models to post-process the output of discriminative models and enhance the separated signals, yielding improvements over the original models. Despite these advances, most approaches still perform significantly lower than discriminative models in objective metrics (SDR).

In speech enhancement, diffusion-based methods have not only achieved comparable performance with discriminative baselines \cite{richter2023speech} but also shown greater robustness under train–test mismatches \cite{lu2022conditional}. Given the similarity between speech enhancement and music source separation, these results motivate further investigation to realize the full potential of generative methods for music source separation.

A straightforward approach to improving generative-based source separation is to train a state-of-the-art (SOTA) diffusion-based speech enhancement model, such as SGMSE~\cite{richter2023speech, richter2025investigating}, on music source separation data, as proposed by Bereuter et~al.\  \cite{bereuter2025towards}. Their model obtained \emph{degradation mean opinion scores} (DMOS) comparable to SOTA Mel-RoFormer, but still performed worse in terms of SDR. Based on the success of the SGMSE model in \cite{bereuter2025towards}, we further investigate diffusion-based source separation in the spectrograms domain. Instead of SGMSE, we apply the EDM diffusion framework proposed by Karras et~al.\ \cite{Karras2022edm}. This choice was made based on the capabilities of the framework to use different model architectures while achieving better performance on image generation with fewer inference sampling steps. Moreover, inspired by Hung et~al.\ \cite{moises-light}, we re-design the DDPM++ model \cite{song2021scorebased} used by EDM~\cite{Karras2022edm} by including band-splitting and dual-path RoFormer blocks. With proper input design, we show that a generative-based approach can achieve comparable SDR scores to discriminative models while outperforming most of the SOTA models on subjective evaluation.

In summary, the contributions of this work are threefold:
\begin{itemize}
  \item We are the first to apply the EDM framework on the vocals source separation task using complex spectrograms as audio representation. This reduces the number of sampling steps compared to previous works (to fewer than 10 steps).
  \item We improved the widely-used DDPM++ model with music-informed design choices specifically for music source separation.
  \item We demonstrate that the proposed generative-based method is capable of achieving competitive objective and subjective scores compared to SOTA discriminative methods.
\end{itemize}

\vspace{-0.1cm}

\section{Background}
\label{sec:background}
Elucidated Diffusion Models (EDM) are trained via denoising score matching \cite{vincent2011connection}, aiming to model the score of a noise-level-dependent marginal distribution of the training data corrupted by Gaussian noise. EDM improves on prior diffusion works in three respects. 

First, it provides a simplified, unified framework whose training and inference are largely model‑agnostic. In other words, the EDM framework is robust to architectural modifications, allowing us to incorporate music‑specific enhancements on top of the DDPM++ model \cite{song2021scorebased}. We will discuss these enhancements in Section~\ref{sec: model}.

Second, EDM optimizes the sampling process to achieve higher quality with fewer steps. Is uses Heun’s second‑order method \cite{ascher1998computer} as ODE solver together with a custom inference noise schedule:

\begin{equation}
    \sigma_{i<N} = \left( {\sigma_{\max}}^{\frac{1}{\rho}} + \frac{i}{N-1}\left( {\sigma_{\min}}^{1/\rho} - {\sigma_{\max}}^{1/\rho} \right) \right)^{\rho}
\end{equation}

${\sigma_{\max}}$ is set to 80 and ${\sigma_{\min}}$ is set to 0.002. Since truncation errors near ${\sigma_{\min}}$ are more impactful, ${\rho}$ is adjusted between 5 to 10 to concentrate (i.e. shorten) steps close to ${\sigma_{\min}}$. These choices allow us to reduce the number of sampling steps and improve inference speed for music source separation, addressing a common limitation of prior generative approaches.

Third, EDM improves training by preconditioning the network with a $\sigma$-dependent skip connection:

\begin{equation}
    D_{\theta}(\boldsymbol{x}; \sigma) = c_{\text{skip}}(\sigma)\,\boldsymbol{x} + c_{\text{out}}(\sigma)\,F_{\theta}\!\left( c_{\text{in}}(\sigma)\,\boldsymbol{x};\, c_{\text{noise}}(\sigma) \right)
\end{equation}

where $F_{\theta}$ is the trained model, $c_{\text{skip}}(\sigma)$ modulates the skip connection, $c_{\text{in}}(\sigma)$ and $c_{\text{out}}(\sigma)$ scale the input and output magnitudes, and $c_{\text{noise}}(\sigma) $ maps noise level $\sigma$ into a conditioning input for $F_{\theta}$. The functions for $c_{\text{in}}$ and $c_{\text{out}}$ are chosen so that network inputs and training targets have unit variance, $c_{\text{skip}}$ so that errors in $F_{\theta}$ are amplified as little as possible, and $c_{\text{noise}}$ is chosen empirically.

%
Finally, because training at intermediate noise levels is most effective, EDM biases training toward the relevant range by sampling $\sigma$ from a log-normal distribution. These modifications speed up training and lead improve output sample quality. For a comprehensive discussion, please refer to the original paper \cite{Karras2022edm}.

\vspace{-0.1cm}

\section{Proposed Method}
\label{sec:proposed_method}

In this section, we describe our adaptations of the EDM framework for music source separation, including changes to the input representation and to the model architecture.

\subsection{Input Representation}
\label{sec: io}

We first compute the complex Short-Time Fourier Transform (STFT) of the stereo mixture and split the real and imaginary parts into separate channels. This results in an image‑like representation, a 3D tensor $C\times F\times T$ with $C=4$, where $F$ and $T$ denote the frequency and time axes.

There are two ways to inject the mixture spectrogram into the diffusion model. The first method is to treat the mixture as a conditioning signal and use the EDM conditioning scheme. However, unlike class or text conditioning, the mixture is a 3D tensor. It requires architectural changes or some kind of preprocessing to use it within the EDM framework. The second method, in contrast, is more straightforward. The mixture spectrogram is concatenated with the noisy target spectrogram along the channel dimension at the input. This simple strategy has shown promising results in speech enhancement \cite{welker22speech} and music generation \cite{nistal2024diff}. We use this concatenation strategy, and expand the input to $C=8$ channels.

Unlike images, spectrograms exhibit a nonuniform energy distribution. Low frequencies typically have higher energy compared to high frequencies, and overall loudness varies across samples. To mitigate this, we first peak-normalize the input waveform $x$ via $x = x / \max(|x|)$. We then apply an amplitude transformation \cite{braun2021consolidated} to the complex spectrogram to enhance lower-energy frequency components:

\begin{equation}
    \tilde{c} = {\beta}|c|^{\alpha} e^{\mathrm{i}\,\angle(c)} \;\Longleftrightarrow\; c = (\frac{|\tilde{c}|}{\beta})^{1/\alpha} e^{\mathrm{i}\,\angle(\tilde{c})},
\end{equation}
where we set $\alpha=0.667$ and $\beta=0.065$ determined via preliminary tuning. 

Prior work \cite{moises-light} has shown that the U‑Net architecture, which is the backbone of the DDPM++ model, benefits from input band‑splitting, allowing the network to process low and high frequencies separately. This is because their energy distributions differ slightly even after normalization. Following these insights, we apply the same efficient band split with number of splits $N_s=4$ to the input spectrogram, yielding $C=32$ and reducing $F$ to $F/4$.

\subsection{Architecture Improvement}
\label{sec: model}

Our model builds on the DDPM++ architecture reimplemented by the EDM authors\footnote{\url{https://github.com/NVlabs/edm/tree/main}}. The model is an encoder–decoder U‑Net with skip connections between corresponding resolution levels. The encoder and decoder each comprise $L$ levels. At each level, they apply a downsampling (encoder) or upsampling (decoder) U‑Net block followed by $N_r$ residual U‑Net blocks, each optionally equipped with a self‑attention layer. The noise level conditioning is implemented using a sinusoidal positional embedding. After passing the noise-embedding through linear and SiLU layers, the embedding is added into every U‑Net block. For further architectural details, see \cite{song2021scorebased}.

Several design choices in the DDPM++ architecture are not optimal for music source separation. Unlike images, the time and frequency axes of spectrograms represent different characteristics of the data. The pixel-wise self-attention used in DDPM++ which treats both axes equally, can therefore lead to suboptimal results. To \amy{address} this, we replace the pixel-wise self-attention layers with dual-path RoFormer blocks \cite{lu2024music}, which process time and frequency axes separately. To stabilize training, the RoFormer blocks are initialized using Xavier uniform, FP32 is used to process rotary embeddings, and the GELU \amy{activation} employs the tanh approximation \cite{lee2024etta}.
Prior work has shown that downsampling and upsampling with transposed convolutions with strides will introduce
aliasing artifacts \cite{stollerWaveUNetMultiScale2018}. To mitigate this, we remove time downsampling from the DDPM++ U-Net.

\vspace{-0.1cm}

\section{Experimental Design}
\label{sec:experiment}

\vspace{-0.2cm}

\subsection{Data}

Following prior work, we use MUSDB18‑HQ \cite{musdb18-hq} as our primary benchmark dataset. MUSDB18‑HQ contains 150 full‑length stereo tracks with a fixed split of 86/14/50 for training/validation/test. To assess the scalability of our proposed model, we also include the larger MoisesDB \cite{moisesdb}, consolidating each track into four stems (vocals, bass, drums, other) to match the stems in MUSDB18‑HQ. Both datasets are stereo at 44.1 kHz sampling rate. Vocals stems from both datasets are used as the training target. 

For training, we randomly sample 6-second excerpts and apply the augmentations from \cite{moises-light}: random mixing, random gain adjustment, polarity inversion, pitch shifting, time shifting, and channel flipping. We compute STFTs with a window size of 2048 and a hop length of 1024. For evaluation, tracks are processed in 6-second chunks with 25\% overlap and reconstructed via overlap‑add to ensure continuity.

\subsection{Hyperparameters}

For our improved DDPM++ variant, we use 128 model channels and $L=4$ levels with channel multipliers [1,2,2,2]. Each level contains $N_r=4$ residual U‑Net blocks, and the noise‑embedding dimensionality is 1024. We apply self‑attention at the end of every residual U-Net block of the encoder and only at the final level of the decoder.

Training is performed with the Adam optimizer, an initial learning rate of $\num{1e-4}$, and a cosine‑annealing scheduler with 4,000 warm‑up steps and one million total steps. Experiments were run on a single H200‑140G GPU with a batch size of 12 for approximately one week. The model is trained with exponential moving average \cite{lu2024music}, and the final checkpoint is used for evaluation.

For diffusion inference, we adopt the EDM default settings: $\sigma_{min}=0.002$, $\sigma_{max}=80$, $\sigma_{data}=0.5$, $\rho=7$, $P_{mean}=-1.2$ and $P_{std}=1.2$. Section~\ref{sec: result} presents ablation studies showing that alternative parameter choices can improve performance by up to 0.6 dB. We did not use Heun’s second‑order sampler, as it increases computation with negligible gains in our experiments.

\vspace{-0.1cm}

\subsection{Evaluation Metrics}
Experiments are evaluated with two metrics. For comparability with prior work, we report chunk‑level Signal-to-Distortion Ratio (cSDR) \cite{stoter20182018} in dB. cSDR is computed on 1‑second chunks. We take the median over chunks for each track and summarize the dataset by the median across tracks. 

\amy{Following} \cite{bereuter2025towards}, which finds that embedding MSE correlate best with degradation mean opinion scores (DMOS) from human listeners, we report the median MSE in the MERT‑L12 embedding space across tracks as a proxy for perceptual quality. \amy{Audio samples are also provided \footnote{\url{https://amymoises.github.io/diffvs.github.io/}}}

\begin{table}[]
    \centering
    \begin{tabular}{l|c|c|c|c}
         \textbf{Model} & \textbf{Params} & $ \rho $ & \textbf{Steps} & \textbf{cSDR$\uparrow$} \\
    \midrule
         DDPM++ & 63.1 M & 7 & 10 & 8.45 \\
         + norm & 63.1 M & 7 & 10 & 8.62 \\
         + norm + arch & 56.7 M & 7 & 10 & 9.53 \\
    \end{tabular}
    
    \caption{Ablation study on different model architecture.}
    \label{tab:ablation}
\end{table}

\begin{figure}[]
  \centering
  \centerline{\includegraphics[width=\columnwidth]{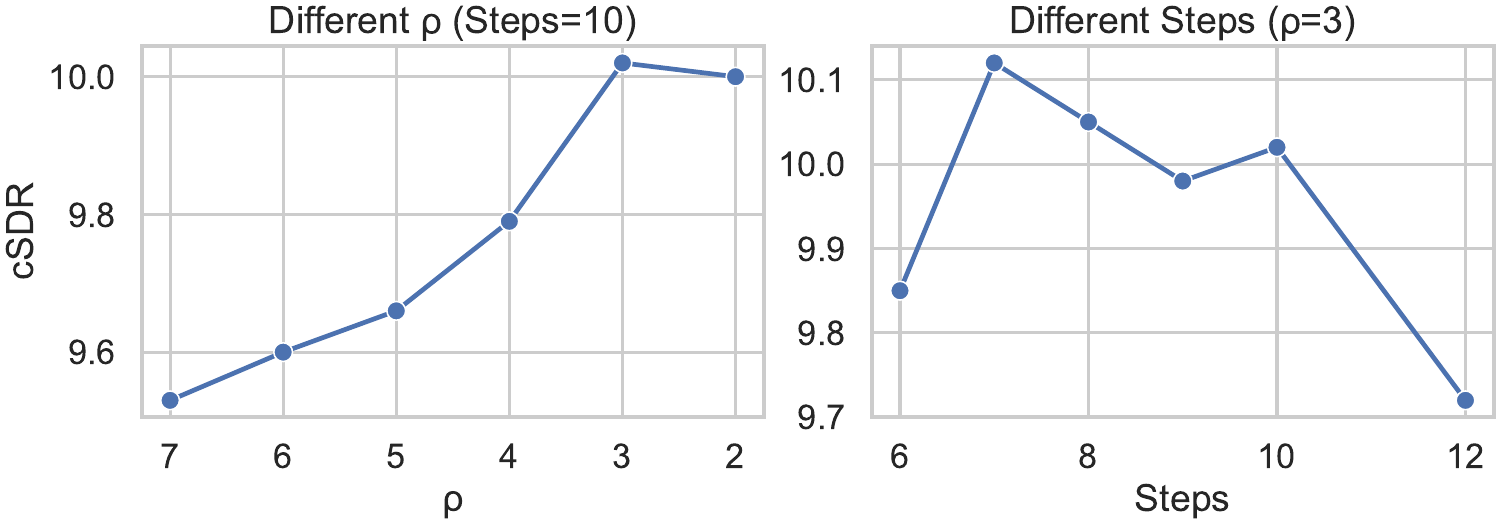}}
  \caption{\amy{Ablation study on different inference parameters.}}
  \label{fig: ablation}
\end{figure}

\section{Results}
\label{sec: result}

Table~\ref{tab:ablation} shows an ablation of the improvements from Section~\ref{sec:proposed_method}. \emph{DDPM++} represents the baseline model using a plain spectrogram as input. \emph{``+ norm''} represents the normalization introduced in Section~\ref{sec: io}. \emph{``+ norm + arch''} represents the additional architectural improvements mentioned in Section~\ref{sec: model}. While both normalization and architectural modifications lead to improvements, the biggest gains stem from architectural changes. In total, SDR improves by over 1 dB compared to the baseline. \amy{Figure~\ref{fig: ablation}} shows results for different inference parameter settings. Within all parameters, we found that adjusting $\rho$ and number of sampling steps has the biggest impact on performance. The middle section of Table~\ref{tab:ablation} shows the effect of different values for $\rho$. Although \cite{Karras2022edm} suggests $\rho=7$ to shorten steps near $\sigma_{min}$ for better performance, we find that smaller values ($\rho=2$ or $\rho=3$), which produce more uniform step sizes, yield better results in our case. The lower part of Table~\ref{tab:ablation} demonstrates the effect of number of sampling steps. The proposed model requires only 7 sampling steps to achieve the highest performance.

\begin{table}[]
    \centering
    \begin{tabular}{l|c|c|c|c}
         \textbf{Model} & \textbf{Type} & \textbf{Params} & \textbf{Extra?} & \textbf{cSDR$\uparrow$}  \\
    \midrule
    \midrule
        HDemucs & disc & 42 M & x & 8.13 \\
        TFC-TDF V3 & disc & 70 M & x & 9.59 \\
        BSRNN & disc & 37 M & x & 10.01 \\
        BS-RoFormer-6L & disc & 72 M & x & 10.66 \\
        SCNet-L & disc & 42 M & x & 10.86 \\
    \midrule
        HTDemucs & disc & 42 M & 800 & 9.20 \\
        BSRNN & disc & 37 M & 1750 & 10.47 \\
        SCNet-L & disc & 42 M & 235 & 11.11 \\
        BS-RoFormer-12L & disc & 93 M & 500 & 12.72 \\
    \midrule
    \midrule
        MSDM & gen & 405 M & x & 3.64 \\
        Diff-DMX-musdb & gen & 99 M & x & 5.38 \\
        SGMSEVS & gen & 65 M & 235 & 8.63 \\
    \midrule
        Ours & gen & 57 M & x & 10.12 \\
        Ours & gen & 57 M & 235 & 10.88 \\
         
    \end{tabular}
    \caption{cSDR Comparison among different source separation models. "disc" stands for discriminative approach while "gen" stands for generative approach.}
    \label{tab:sdr}
\end{table}

Table \ref{tab:sdr} reports cSDR results compared to SOTA baselines, which are trained with the same train/validation/test splits from MUSDB18-HQ, with or without additional training data. We evaluate against five SOTA discriminative models, HDemucs \cite{defossez2021hybrid}, TFC–TDF v3 \cite{kim2023sound}, BSRNN \cite{luo2023music}, BSRoformer \cite{lu2024music}, and SCNet \cite{tong2024scnet}, and three generative models, MSDM \cite{mariani2023multi}, Diff-DMX-musdb, and SGMSE \cite{welker22speech} retrained by \cite{bereuter2025towards} on MUSDB‑HQ and MoisesDB (SGMSEVS) \footnote{The SDR results are calculated from the evaluation CSV files in  \url{https://github.com/pablebe/gensvs_eval/tree/main}}.

Among all discriminative baselines, our model outperforms HDemucs, TFC–TDF v3, and BSRNN, and performs only slightly worse than BSRoformer-6l and SCNet. This indicates that our generative approach attains objective performance comparable to leading discriminative systems, addressing a common weakness of prior generative methods. Compared with other generative approaches, our model surpasses all prior work by roughly 1.5 dB (cSDR), including SGMSE, which is widely regarded as one of the SOTA models in speech enhancement. Note that the version of SGMSE used for this evaluation was trained on more data (MUSDB18-HQ + MoisesDB). Additionally, our model has the fewest parameters among all generative baselines and requires only 7 inference steps, versus 150 for MSDM, 20 for Diff‑DMX‑musdb, and 35 for SGMSEVS.

With additional training data from MoisesDB, our method outperforms HTDemucs and BSRNN while using only 235 extra tracks (vs. 800 for HTDemucs and 1,750 for BSRNN). The SDR score is slightly below SCNet‑L and about 2 dB behind BSRoformer. 


\begin{table}[]
    \centering
    \begin{tabular}{l|c|c|c|c}
        \textbf{Model} & \textbf{Type} & \textbf{Params} & \textbf{Extra?} & \textbf{MSE$\downarrow$} \\
    \midrule
        SCNet-L & disc & 42 M & 235 & 0.096 \\
        Mel-RoFormer & disc &  228 M & unkown & 0.071 \\
        SGMSEVS & gen & 65 M & 235 & 0.089 \\
    \midrule
        Ours & \amy{gen} & 57 M & 235 & 0.083 \\
    
    \end{tabular}
    \caption{Compared with different SOTA models with MERT embedding MSE.}
    \label{tab:mse}
\end{table}

Table~\ref{tab:mse} presents the subjective evaluation using embedding MSE as a proxy. We compare three SOTA separation models: SCNet‑L (open‑source SOTA discriminative model), Mel‑RoFormer (literature SOTA discriminative model, retrained by open-source project \footnote{\url{https://github.com/KimberleyJensen/Mel-BandRoformer-Vocal-Model}}, and SGMSEVS (SOTA generative model). Although our model’s SDR is lower than SCNet‑L, it achieves better embedding MSE than both SCNet‑L and SGMSEVS, indicating a perceptual advantage over the baseline models of similar size.

\section{Conclusion}
In this paper, we adapt the EDM framework for vocal source separation. Through architectural improvements, data normalization, and inference parameter optimization, our model outperforms prior generative systems and achieves SDRs comparable to strong discriminative baselines. For future work, we plan to extend the approach to other instruments, including the standard four stems, and to investigate diffusion modeling for music source separation more deeply, exploring potential improvements to both the network architecture and input representation.

\bibliographystyle{IEEEbib}
\bibliography{strings,refs}

\end{document}